# Protecting and Promoting Human Agency in Education in the Age of Artificial Intelligence

**Authors and Affiliations:**
Olga Viberg*, KTH, Stockholm, Sweden
Mutlu Cukurova*, University College London, London, United Kingdom
Rene F. Kizilcec*, Cornell University, Ithaca, NY, United States

Simon Buckingham Shum, University of Technology Sydney, Australia
Dorottya Demszky, Stanford University, Stanford, CA, United States
Dragan Gašević, Centre for Learning Analytics, Faculty of Information Technology, Monash University
Thorben Jansen, Leibniz Institute for Science and Mathematics Education, Kiel, Germany
Ioana Jivet, CATALPA, FernUniversität in Hagen, Germany
Jelena Jovanovic, University of Belgrade, Serbia
Jennifer Meyer, University of Vienna, Vienna, Austria
Kou Murayama, Hector Research Institute of Education Sciences and Psychology, Germany
Zach Pardos, UC Berkeley, United States
Chris Piech, Stanford University, Stanford, CA, United States
Nikol Rummel, Ruhr-University Bochum and CAIS, Bochum, Germany
Naomi E. Winstone, Surrey Institute of Education, University of Surrey, United Kingdom

* Equal contribution

**Corresponding authors:**
O. Viberg (oviberg@kth.se), M. Cukurova (m.cukurova@ucl.ac.uk), R. F. Kizilcec (kizilcec@cornell.edu)

# Protecting and Promoting Human Agency in Education in the Age of Artificial Intelligence

**Abstract**
Human agency is crucial in education and increasingly challenged by the use of generative AI. This meeting report synthesizes interdisciplinary insights and conceptualizes four aspects that delineate human agency: human oversight, AI-human complementarity, AI competencies, and relational emergence. We explore practical dilemmas for protecting and promoting agency, focusing on normative constraints, transparency, and cognitive offloading, and highlight key tensions and implications to inform ethical and effective AI integration in education.

**Introduction**
Understanding how human agency is sustained and develops through interactions with artificial intelligence (AI) in education is crucial for safeguarding individual autonomy and guiding the responsible and effective integration of technology. In June 2025, researchers from multiple disciplines gathered at the "Fostering Human Agency in AI-powered Education" workshop in Marbach Castle, Germany, to address theoretical, empirical, and practical challenges of maintaining and promoting human agency amidst growing reliance on AI. Human agency, defined as the capacity to act intentionally, make informed choices, exert meaningful control, and influence the course of events, is central to educational practice due to its ethical and practical implications [1-3]. Ethically, education is a humanistic pursuit fostering empathy, tolerance, and intercultural understanding, recognized in the UN Universal Declaration of Human Rights [4]. Practically, scholars warn that diminishing human agency may erode cognitive, metacognitive, and affective skills essential for lifelong learning [5-7]. In many school systems, agency is constrained by rigid curricula, high-stakes testing, and limited influence over what and how people teach and learn. This sociopolitical reality underscores the need to deploy AI in ways that enhance rather than undermine human agency, ensuring that teachers and learners remain active decision-makers capable of interpreting, adapting, or overriding AI-generated outputs [8].

Human agency in education has been examined through various theoretical lenses, highlighting its role in learners' and educators' decision-making and autonomy [9]. Theoretical perspectives, including sociocultural [10], self-determination [3,11], and social cognitive theories [12], characterize agency as relational, context-sensitive, and shaped by autonomy, social interaction, and self-regulation. The workshop called for an interdisciplinary framework integrating cognitive, social, educational, and computational insights with practical design principles for AI-enhanced education. While participants noted advances in technologies to explore human agency in AI-mediated contexts, they also recognized persistent challenges in harmonizing methodologies and concepts across disciplines.

This report aims to foster continued interdisciplinary collaboration by synthesizing multidisciplinary perspectives on human agency in education. First, we integrate and refine conceptualizations of four sources of agency: *human oversight*, *AI-human complementarity*, *AI competencies*, and *relational emergence*. Then, we examine four practical dilemmas in aligning AI technologies with human values and educational goals, highlighting key tensions and practical implications to help researchers and practitioners design and evaluate AI systems that genuinely support human agency in AI-mediated educational settings.

**Conceptualizing Sources of Human Agency with AI in Education**

Human agency is subject to influence by personal, social, contextual, and technological factors, including power dynamics, historical contexts, and characteristics of AI-mediated socio-technical systems [13]. Although AI can enhance agency by empowering learners and educators in accomplishing challenging tasks, it can also undermine their agency: for instance, by causing over-reliance on AI tools, loss of meaningful control (e.g., individuals' voice and intellectual contributions in writing), and reduced decision-making opportunities for learners and educators [14]. Workshop participants identified seminal work on human agency, which we synthesized into four conceptualizations that provide complementary perspectives on how to protect and promote agency around AI. These conceptualizations were explored in structured scenario-based group discussions, where multidisciplinary teams engaged with concrete challenges involving three specific AI systems.

*Human Oversight of AI*

Human agency can be safeguarded in educational contexts through mechanisms that ensure AI systems remain under "meaningful human oversight," as proposed by Cavalante et al. [15]. The authors outline four conditions for maintaining moral and operational authority over AI: (1) bounding AI's operational domain in morally sensitive areas, (2) ensuring humans and AI share compatible task representations, (3) aligning human responsibilities with actual capacity for control, and (4) creating clear links between AI actions and human decisions. As AI grows increasingly autonomous, oversight becomes more pressing, requiring thoughtful distribution of control where AI actions remain subordinate to human intentions (e.g., human-in-the-loop designs; [16,17]). However, rapid advances in generative AI blur oversight boundaries and expand AI's capacity for complex tasks.

Tensions arise between maintaining meaningful human oversight and institutional pressures for efficiency and cost reduction, which may incentivize full automation at the expense of human decision-making. Moreover, effective oversight of AI often involves monitoring system behaviour through extensive data collection, raising concerns about privacy and potential misuse of educational data by technology companies [18]. Governance-oriented frameworks [19, 20] address these concerns by emphasizing the need for ethical design, transparency, accountability, and clear normative guidelines. Floridi's [20] concept of "agency without intelligence" emphasizes that as AI systems become more autonomous regardless of their "intelligence" they still require careful human oversight.

Participants noted that, while oversight can promote agency by providing scaffolding, transparency, and accountability, it may also constrain agency if it results in excessive monitoring, rigid frameworks, or diminished autonomy. There was consensus that no single form of oversight is sufficient: effective agency-supporting oversight demands diverse, coordinated expertise that is pedagogical, technical, content-specific, and practical. Participants warned that treating oversight as a universal solution overlooks deeper systemic issues [21]. They called for moving beyond simplistic control models, recognizing oversight as a necessary but insufficient condition for sustaining meaningful human agency.

*Human-AI Complementarity*

Human agency can be safeguarded by designing AI systems to complement rather than replace human capabilities. Rather than viewing AI merely as a tool, emerging perspectives frame human–AI interaction as a partnership in which each side's strengths mitigate the other's limitations. Holstein et al. [22] describe this as *Hybrid Adaptivity*, where educators and AI systems co-construct adaptivity across perception, interpretation, decision, and action. Sundar [23] similarly argues that agency is strengthened when users can customize AI recommendations. These perspectives converge on the view of AI as an assistive agent, supporting high-order human thinking and ethical decision-making, as well as emotional companions for students and teachers [24].

Participants emphasized that meaningful complementarity requires intentional design, not just division of labor. As AI systems grow more complex, it becomes harder for users to understand or adjust outputs. Feedback loops that preserve explainability and interpretability are critical to ensure alignment with human goals. This echoes long-standing concerns from educational psychology that overly directive teaching can reduce student agency [25-27]. This is a tension mirrored in concerns that AI may undermine learner and teacher autonomy. For instance, while Intelligent Tutoring Systems (ITS) can aid students, they risk reducing teacher agency when educators are excluded [28]. New tools like PromptHive [29] seek to counter this by involving educators in co-authoring AI-generated feedback, reinforcing professional judgment. Similarly, offloading metacognitive tasks to AI may streamline learning but risk reducing students' ownership over their learning process [30].

Cognitive science perspectives further illuminate the complexities of human-AI complementarity. Chiriatti et al. [31] introduce "System 0," describing AI as a third cognitive layer external to humans, capable of processing information and producing outputs that users may accept without deep scrutiny. In dual-process theories of cognition, System 1 refers to fast, intuitive, automatic thinking, while System 2 denotes slower, deliberative, and analytical reasoning [32]. While AI extends human cognitive capabilities, it risks promoting machine-like reasoning habits, such as treating statistical correlations as truths and bypassing critical interpretation. Participants debated whether agency is a zero-sum game (i.e., whether gains in efficiency from AI necessarily diminish human reflection).

Participants emphasized that agency is not granted or withdrawn but co-constructed through iterative design. They advocated gradual AI integration aligned with co-adaptive strategies that keep agency distributed and context-sensitive. Much of the discussion focused on how to maintain pedagogical rigor by supporting, rather than displacing, teacher agency. For example, design strategies such as delegating lower-level tasks to AI while preserving teacher control over complex decisions, or generating multiple options to prompt pedagogical reflection [33], were discussed as ways to preserve meaningful human engagement.

Finally, participants underscored that human interaction remains irreplaceable in education. Effective designs might initially constrain AI's role, gradually expanding its capabilities as learners and educators gain skills and confidence, echoing scaffolding approaches from learning sciences (e.g., [34,35]). Participants called for rigorous empirical studies to identify how best to configure human-AI partnerships that preserve agency across professional, social, cognitive, and pedagogical domains.

*AI Competencies*

Human agency can be safeguarded by empowering users of AI through education and literacy, enabling them to engage critically with AI systems and their outputs. Since it is impossible to design AI that is transparent and explainable to users with varying literacy levels, educational stakeholders need to learn how to navigate AI's inherent opaqueness, biases, and partiality. Bearman and Ajjawi [36] advocate for pedagogy that helps learners work productively with the "black box" nature of AI, emphasizing resilience in ambiguity and cultivating judgment about how to act on AI output. Within this view, competencies such as critical thinking, self-directed learning, creative problem-solving, and strategic decision-making are positioned as central to preserving human agency. While these are not universally defined as core AI competencies, they are frequently cited in AI literacy approaches [37] as essential dispositions for navigating AI-mediated environments. To avoid conflating broader educational goals with AI-specific competencies, it is essential to clearly articulate how and why these particular skills are emphasized in this context. Developing them requires not only teaching of AI concepts and ethics but also hands-on, constructivist experiences that enable learners and teachers to become reflective agents rather than passive recipients of algorithmic outputs.

Participants stressed that AI competence extends beyond understanding AI's mechanics; it includes knowing when to trust, challenge, or override AI outputs and maintaining critical distance to preserve human judgment, especially in teaching and learning (cf. [38]). Authentic practice in safe and low-stakes environments was seen as essential for building these capabilities, especially given that AI's ease of use may reduce critical thinking [39] and promote (meta)cognitive laziness [5]. Importantly, learners and teachers also need to be motivated to utilize these competencies: Demszky et al. [40] found that offering educators agency in AI-mediated professional learning was most beneficial to those educators who were already motivated to self-improve. Several participants emphasized that motivation is essential for sustaining human agency and applying relevant competencies. Individuals may relinquish agency by passively accepting AI outputs, such as submitting unexamined AI-generated work or teaching uncritically from an AI-generated lesson plan. Cultivating both motivation and critical competencies is therefore vital to ensure that learners and educators actively exercise agency alongside AI.

Workshop discussions also highlighted that thriving with AI requires uniquely human knowledge work and critical engagement. Learning perspectives stress resilience, AI literacy, and ethics [36,41]. Participants linked AI competencies to self-regulated learning, emphasizing that learners must treat AI as a strategic aid, not a crutch. Differentiating performance with AI from authentic learning is key [42], calling for educational practices that scaffold agency. Training should support discerning when to trust or challenge AI while maintaining agency.

*Emergent Relational Design*

Human agency can be safeguarded through the collective design of systems and practices that align AI capabilities with human intentions and values, considering that agency in human-AI systems emerges *relationally* across sociotechnical networks and may not be located solely within the human or the machine. The *Postdigital Learner Agency* framework [43] conceptualizes agency as adaptive, co-constructed, and embedded, shaped by interactions between people, technologies, and institutional structures. Drawing on sociocultural theory and postdigital philosophy, this perspective shifts the focus from individual autonomy to *distributed agency*, which is context-sensitive and contingent on available tools, routines, and norms. This ecological perspective is echoed by [44], who view human capability as emerging from within complex, interconnected learning ecologies. Rather than isolating humans from AI to protect agency, these perspectives argue for managing entanglements and designing systems that reinforce human intentionality and control.

Participants emphasized that agency is embedded in social, institutional, and technological contexts, requiring individuals to negotiate and navigate collective practices and institutional support alongside AI. Participants discussed *systemic agency* as extending beyond individuals to include actors such as parents, peers, and institutions: for example, parental agency might diminish if AI assumes roles like homework assistance, prompting questions about appropriate counterfactual experiences and how human roles should be redefined. Hybrid intelligence frameworks (e.g., [30,45]) highlight that effective AI integration depends on intentional co-design and context-sensitive oversight.

Participants raised several concerns in the context of systemic agency. First, AI could subtly homogenize and standardize cognitive and social experiences [46], potentially narrowing diversity in how individuals think and interact. Second, academic integrity boundaries are challenged by the increasingly sophisticated outputs that AI produces, raising questions about who determines plagiarism or acceptable similarity, and how to design assessments to constrain GenAI use [47]. Third, how to best redeploy human time saved by AI automation is unclear. For instance, if AI generates feedback, educators could allocate time to helping students engage meaningfully with that feedback or divert efforts to other tasks (e.g., [48]). This illustrates that AI-based automation (even without human oversight for certain tasks) has the potential to increase human agency by enabling redistribution of work in ways that increase agency beyond merely spending effort on human oversight of AI.

**Core Dilemmas of Human Agency with AI in Education**

To connect theory with practice, the workshop included debates on critical dilemmas concerning AI and human agency, grounded in the four conceptualizations discussed earlier. Participants examined tensions and philosophical questions through three key dilemmas: (1) Should AI be normatively constrained to support human agency? (2) Should people be educated to cope with black boxes to protect and promote their agency, or should the technology be made to be explainable and transparent? (3) If AI is System 0, what kind of cognition is it, and what are the implications of this on the allocation of decision-making tasks? We synthesize core arguments and insights articulated for each dilemma.

*Normative Constraints on AI to Optimize Levels of Human Agency*

Participants debated whether AI should be normatively constrained, recognizing that, unlike simple tools, AI exhibits complex behaviors that influence human action. Since AI is not a moral agent, responsibility for embedding ethical "ought-to-be" norms falls on designers, developers, and educators. However, defining these norms raises challenges: whose values should prevail, and could rigid constraints unintentionally standardize education or deepen inequities through filter bubbles (e.g., [14,44])? Participants noted the practical difficulty of aligning evolving AI technologies with normative constraints [17], amid commercial pressures.

While immediate solutions are limited, participants saw value in theory- and evidence-based benchmarks to assess AI's pedagogical impact on agency [19,20]. Responsible use statements were suggested as a bottom-up way to align tools with stakeholders' values. Preserving human agency requires both technical design and societal dialogue, ensuring individuals feel empowered without being overburdened [13,21].

*Educating People to Cope with Black Boxes vs. Demanding Explainable AI*

The debate centered on whether protecting human agency demands striving for inherently explainable systems or educating users to navigate AI's opacity. Participants noted that, while fully explainable AI could be ideal, explainability in complex AI systems is challenging to achieve, and transparency does not always translate into user comprehension [17]. Moreover, humans and human learning themselves can be considered "black boxes," suggesting that AI transparency should be purpose-driven to enable oversight and informed decisions rather than transparency for its own sake.

Given these challenges, participants agreed that modern learners must understand how algorithms influence choices and information access, echoing Bearman and Ajjawi's [36] call for resilience in ambiguity rather than reliance on technical explainability alone. A dual approach emerged as the likely path forward: continue technological efforts to improve transparency where possible while simultaneously preparing users to operate effectively amidst inevitable ambiguity. The level of explainability required depends on the user and context, underscoring the need for contextual, user-centered AI design in education [22]. In tools like ITS, for example, teachers should clearly understand how AI generates feedback and whether their materials are used, maintaining professional and pedagogical agency.

*AI as System 0 and Implications for Human Cognition and Decision-Making*

Participants explored whether AI constitutes a new cognitive system (*System 0*) as proposed by Chiriatti et al. [31]. They noted that tools like search engines, recommendation systems, and generative AI involve outsourcing tasks such as memory, pattern recognition, and creative thinking, aligning with theories of extended cognition [49]. While this offloading frees cognitive capacity for higher-order reasoning, it also risks making human cognition interdependent with technology, leaving it vulnerable to manipulation by developers' design choices.

Key concerns focused on managing the interplay between AI (System 0) and human cognitive systems (Systems 1 and 2). Participants cautioned that AI biases could influence human biases, necessitating deliberate, reflective reasoning to evaluate AI outputs critically. This connects to the notions of complementarity and competencies: although AI can support human thinking, individuals need skills to decide when to trust AI and when to intervene. Even choosing to use AI (or not) requires reflective engagement, but sustaining critical scrutiny of AI's outputs remains a challenge. Participants concluded that while AI can enhance human cognitive capabilities, careful attention is needed to ensure it does not erode human autonomy, freedom, or the social fabric of democratic institutions.

**Reflection and Future Directions**

Safeguarding human agency around AI in education requires navigating complex interdisciplinary tensions. Many critical questions remain unresolved, and answers will depend heavily on the specific learning context: How can we balance human oversight with institutional pressures for efficiency and automation? What competencies do learners and educators need to engage critically with AI, and how should these be developed? How can systems be designed to support distributed, relational agency while avoiding homogenization or inequity? Where should boundaries be drawn between tasks best delegated to AI and those requiring human judgment? Beyond safeguarding agency, we note that AI can also increase human agency when used to improve education, for instance strengthening learners' competencies through feedback and other pedagogically sound practices.

Addressing these questions and opportunities will require collaboration across research, policy, and practice. Researchers should prioritize longitudinal and context-specific studies to identify effective designs and evaluations for human-AI partnerships, while also advancing theory to deepen our understanding of human agency in AI-mediated learning. Policymakers must create regulatory frameworks that protect human agency without stifling innovation. Educators and developers should co-design tools that enhance, rather than undermine, human judgment and creativity. Ensuring AI amplifies human dimensions of education requires dialogue, evidence, intentional design, and firm theoretical grounding.

**Acknowledgments**

The workshop was supported by STINT, the Swedish Foundation for International Cooperation in Research and Higher Education (grant agreement number MG2018-7984), and the Jacobs Foundation.

**Author contributions**

O.V., M.C., and R.F.K. organized the meeting and wrote the first version of the report. All authors contributed to the revision of the manuscript, and read and approved the final version.

**Competing Interests**

We declare no competing interests.